# A silicone-based slippery polymer coating with humidity-dependent nanoscale topography.

M. Callau,[a] C. Fajolles,[a] J. Leroy,[b] E. Verneuil,[c] P. Guenoun [a]*

[a] Université Paris-Saclay, CEA, CNRS, NIMBE UMR 3685, LIONS, 91190 Gif-sur-Yvette, France

[b] Université Paris-Saclay, CEA, CNRS, NIMBE UMR 3685, LICSEN, 91190 Gif-sur-Yvette, France

[c] Soft Matter Sciences and Engineering (SIMM), ESPCI Paris, PSL University, Sorbonne Université, CNRS, F-75005 Paris, France

*Corresponding author: Dr. P. Guenoun

    NIMBE, UMR 3685 CEA, CNRS
    Université Paris-Saclay
    CEA Saclay 91191 Gif-sur-Yvette Cedex (France)
    Telephone: 06.84.22.67.22
    E-mail: patrick.guenoun@cea.fr

**Abstract**

*Hypothesis:* Slippery Omniphobic Covalently Attached Liquids (SOCAL) have been proposed for making omnirepellent thin films of self-assembled dimethylsiloxane polymer brushes grafted from silica surfaces. Smooth and flat at very small scale, these fluid surfaces could exhibit a more complex multiscale structure though showing very weak contact angle hysteresis (less than 5°).

*Experiment:* In this work, coatings were deposited on glass surfaces from an acidic dimethoxydimethylsilane solution under carefully controlled relative humidity. Ellipsometry mapping was used to analyze the surface structuration with nanometric thickness sensitivity. The sliding properties were determined using a drop shape analyzer with a tilting device, and chemical analyses of the coatings were performed using on-surface techniques (XPS, ATR-FTIR spectroscopy).

*Findings:* Coated materials possessed an unexpected surface structure with multiscale semispherical-like features, which surprisingly, did not increase the contact angle hysteresis. A careful study of some parameters of the coating process and the related evolution of the surface properties allowed us to propose a new model of the chemical organization of the polymer to support this remarkable liquid-like behavior. These structures are made of end-grafted strongly adsorbed Guiselin brushes with humidity-dependent thickness: the higher the humidity, the thinner and the more slippery the coating.

**Graphical abstract**

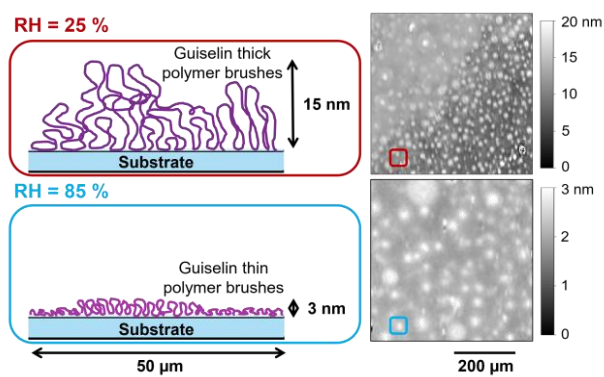

# 1. Introduction

Omniphobic, or more exactly, omnirepellent coatings are highly requested for protecting all kinds of surfaces such as glass [1–4], metals [5,6] or textiles [7,8] as they are expected to provide self-cleaning [1,9–11] and anti-fogging abilities [12,13]. The basic property is that liquid drops of either water solutions or water-immiscible oil-like compounds slide easily on the surface under a small inclination or force.

A first approach to omnirepellency is to maximize the contact angle of a drop on a substrate, the adhesion forces being minimized as the contact surface between the liquid (drop) and the solid (substrate) is reduced, facilitating the sliding of the drop. In 1941, polytetrafluoroethylene (PTFE) was accidentally discovered by Plunkett [14]. Better known as Teflon, this fluoropolymer possesses surprisingly slippery and non-stick properties, induced by its low surface energy [15–18]. It opened the way to the use of fluorinated surfaces where the contact angle of both oils and water can be surprisingly low [19]. Nevertheless, recent studies have classified the perfluorooctanoic acid (PFOA), a by-product of the fluoropolymer industry, as a persistent, bio-accumulative, toxic, and potentially carcinogenic organic product for humans [20–22]. It is therefore urgent to find safer approaches to fluorinated surfaces. One alternative can be inspired by nature and has stemmed biomimetic structures. This physical or chemical surface modification consists in structuring a surface of interest (often composed of a hydrophobic material) at the nanometric, micrometric or both scales, reproducing the structuration found on animal skins, insects, or plants [23–26]. This structuring allows to create a gaseous interface between the surface and the liquid. The latter then rests on a solid/air composite surface which eases the sliding of liquid droplets. However, structured surfaces are very fragile and sensitive to pressure.

However, and especially for oils, the contact angle does not have to be larger than 90° to exhibit the easy-sliding property. Indeed, if the contact angle hysteresis is null, even drops of low contact angle easily slide. Therefore, omnirepellent surfaces that allow an easy sliding of droplets do not necessarily have to be omniphobic surfaces (contact angles larger than 90° for water and oil droplets).

In recent years, a very promising method to nullify the hysteresis was designed by making the drop to be effectively in contact with a liquid coating at the surface of the substrate. A first method was proposed by Quéré *et al.* in 2005 from a theoretical point of view [27]. Inspired by the peristome of carnivorous plants "Nepenthes", it has later been experimentally made by Aizenberg's group by trapping a sparsely soluble oil in a nanostructure [28]. Though very promising, this method needs the nanostructure and the oil to be stable on the long run, a practically difficult task.

Therefore, siloxanes (mostly polydimethylsiloxane (PDMS) molecules) seem to be promising candidates for the creation of omnirepellent coatings, as the high mobility of the Si-O-Si bond of these molecules allow them to form liquid-like surfaces [29], minimizing contact angle hysteresis. The grafting of such a nanometric and soft siloxane-like layer on substrates was pioneered by McCarthy's group in 2016 [2]. These surfaces are usually made through a grafting-from mechanism [30] with siloxane monomers, making PDMS brushes. The coating surfaces are then usually called Slippery Omniphobic Covalently Attached Liquids (SOCAL) as only one side of the molecules (polymer chains) is supposedly covalently grafted to the substrate. The impact of the molecular weight, the molecular structure, or the grafting density of PDMS on the properties of liquid-like coatings (thickness, sliding behavior) have been well described in a recent review [29].

However, it is always assumed that PDMS molecules are in a brush-like configuration, but other molecular organizations could also be possible (loop-like configuration [31], adsorption [32–34]). In addition, in the process conditions used by McCarthy *et al.* (acidic solution with 65 % humidity [2]), the hydrolysis and polymerization of siloxanes is fast [35] such that a pure grafting-from mechanism might not be relevant. The topography of the surface must also be carefully determined as surface structuration might also lead to slippery surfaces [36–39]. These latter aspects were at the origin of our questioning whose results are presented here.

In this work, coatings are deposited on glass slides from a solution composed of dimethoxydimethylsilane (DMDMOS), 2-propanol and sulfuric acid, through a specific dip-coating process, adapted from the one presented by McCarthy *et al.*'s group [2]. Surface topography is examined quantitatively by multiscale methods, namely ellipsometry mapping for large lateral scales and nanometric thicknesses, and atomic force microscopy for smaller scales both laterally and in the z direction. All these methods concur for evidencing a multiscale structuration at the surface with nanometric thickness and micrometric width, highly sensitive to humidity during the condensation step of molecules. Surprisingly enough, in some case, this surface structuration does not preclude the needed reduction of hysteresis for either water, or oil drops. An alternative molecular organization to the one proposed by McCarthy *et al.*'s group is presented. This organization depends on the relative humidity used during the condensation of siloxane molecules. A link between the molecular organization, the surface structuration and the slippery behavior of the coating is evidenced, highlighting the beneficial role of adsorbed species in reducing hysteresis.

## 2. Materials and methods

### 2.1. Materials

2-propanol (iPrOH, > 99.8 %), was purchased from Honeywell (Germany). Sulfuric acid ($H_2SO_4$, > 96 %), Dimethoxydimethylsilane (DMDMOS, > 95 %) and Hexamethyldisilazane (HMDS, > 98 %) were purchased from Thermo Fisher Scientific (France). Toluene (> 99.8 %) was purchased from Carlo Erba (Italy). Dimethylsulfoxide (DMSO, > 99.9 %) and Tetradecane (> 99 %) were purchased from Sigma-Aldrich (Germany). These ingredients were used without any further purification. Deionized water (18.2 MOhm·cm resistivity) was used for all rinsing and wetting processes.

Glass slides (floated soda-lime glass, 76 mm x 26 mm x 1 mm) were purchased from VWR (France). Single side polished silicon wafers (diameter: 4 ", orientation: (111), N-doped) were purchased from Sil'tronix Silicon technologies (France).

### 2.2. Sample preparation

**Coating solution**

The coating solution used in this study was inspired from the process of Wang and McCarthy [2]. Typically, the solution was prepared by mixing, in a glass flask, 427 µL $H_2SO_4$ (1 wt.%) with 8.92 mL DMDMOS (9 wt.%) in 100 mL iPrOH. The solution was then stirred by hand for 30 s and allowed to rest at room temperature.

**Dip-coating process**

Before coating, the glass slides or silicon wafers were cleaned in an acidic solution (10 mL iPrOH and 1 mL $H_2SO_4$) for 60 s and rinsed off thoroughly in water and iPrOH. Right after, the substrate was immerged in the solution with a Riegler&Kirstein dip-coating apparatus at 2.5 mm.s$^{-1}$ speed, kept in the solution for 60 s and withdrawn at 0.1 mm.s$^{-1}$ speed. This withdrawal speed has been chosen to deposit the thinnest coating layer and to improve the homogeneity of the coating [40]. The coated sample was then placed in a closed chamber at room temperature (20 – 25 °C) and under controlled relative humidity (RH) for 30 min, allowing the molecules to interact with the surface of the substrate and leading to the formation of a coating layer (condensation step). To control the relative humidity, saturated solutions of potassium chloride KCl (RH = 85 %), sodium nitrate $NaNO_2$ (RH = 65 %) and silica gel (RH = 8 – 35 %) have been used. Some samples have been prepared at ambient RH (50 %). After this condensation step, the sample was rinsed by immersion in Milli-Q water, in iPrOH and/or in toluene, according to the chosen process (written as $H_2O$+iPrOH and $H_2O$+iPrOH+toluene rinsings).

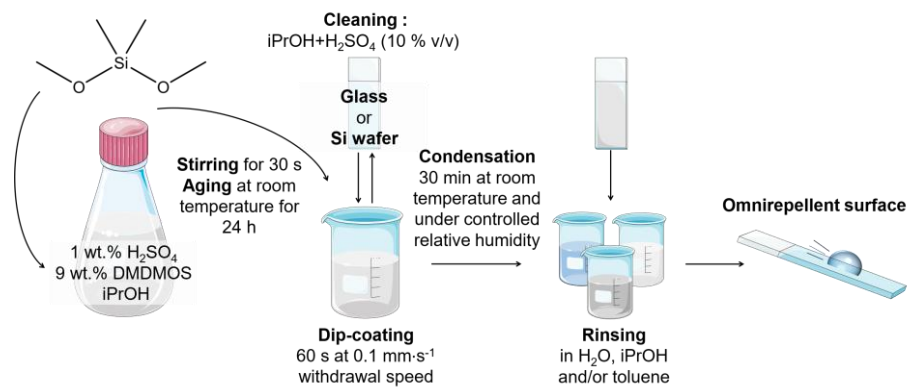

*Figure 1.* Schematic of the dip-coating process of a glass slide in a DMDMOS-based solution (top left molecule).

All the different coating processes were repeated at least two or three times and similar results (sliding behavior, topography, molecular composition) were always obtained.

**HMDS silylation**

To evaluate the presence of hydroxyl groups on the coatings surfaces, HMDS has been used through three processes. For the liquid phase processes: HMDS have been diluted in iPrOH (15 % v/v) [41] or used in pure form (30 mL in a beaker). For the gas phase process, 5 mL of HMDS have been poured into a beaker. The coated glass slides have been placed vertically in all three beakers (immersed for the liquid processes). The beakers were closed hermetically, and the samples have been left to react for 24 hours at room temperature (20 – 25 °C) before analysis.

## 2.3. Surface topography: ellipsometry mapping

Ellipsometry mapping was used to measure the thickness of the grafted layer and to image the surface topography of the sample at large scale (800 µm x 1600 µm).

The measurements were performed on samples prepared on silicon wafers with a Nanofilm EP3 from Accurion, equipped with a 5x objective and a camera. The light source is a Xenon lamp with a filter at wavelength λ = 590 nm. The angle of incidence was fixed at 65°.

For each pixel of area (around 2 µm$^2$), nulling ellipsometry [42] was performed which provides a mapping of the ratio of the global reflection coefficients of polarizations p and s. By comparison with theoretical models computed for a 4-layer system (silicon/silica/polymer/air) [43], the measurements are converted into thickness maps of the polymer layer, provided that refractive indices of each layer are known as well as the silica layer thickness. For each sample, the mean silicon oxide layer thickness is measured on a bare piece of the same wafer. We typically find 2.7 nm. The refractive indices were set as follows: 1 for air; 1.4 for the coating layer (PDMS); 1.46 for silica; 3.94 + 0.02*i* for silicon. We obtained maps for the thickness of the polymer layer as shown in Figure 2.

## 2.4. Wetting properties

The wetting properties of the coatings, namely the static contact angles CA, the hysteresis CAH and the sliding angles SA, have been evaluated with a goniometer.

The droplets analyses were conducted by a DSA25 Drop Shape Analyzer from Krüss Scientific. The apparatus was equipped with a tilting device PA4020 and a camera attached to the tilting device. The angle values were automatically measured by the Advance software, with an internal image treatment tool using a tangent fitting method. To cover a large range of surface tensions, water ($\gamma_{LV,20\,°C}$ = 72.8 mN·m$^{-1}$), DMSO ($\gamma_{LV,20\,°C}$ = 43.5 mN·m$^{-1}$) and tetradecane ($\gamma_{LV,20\,°C}$ = 26.6 mN·m$^{-1}$) were chosen as probe liquids. For CA measurements, a droplet of 2 µL of liquid was placed on the surface and the angle value was recorded for 30 s. For SA measurements, a droplet of 20 µL of liquid was deposited on the surface and the sample was tilted from 0 to 90° at 1°·s$^{-1}$ speed. The SA was determined when both edges of the droplet detached from the surface. CAH was calculated by subtracting the receding angle to the advancing angle, measured for the SA value. A study on the reliability of this method has been carried out and can be found in *Supporting Information*.

Both CA and SA experiments were made at least 3 times on each sample, at different locations and repeated for each one of the probe liquids. The CA, SA and CAH values are averages of these 3 measurements and are reported in degrees (°).

## 2.5. Chemical composition of the on-surface coating

**Attenuated total reflectance Fourier-transform infrared spectroscopy (ATR-FTIR)**

A qualitative detection of PDMS was made by ATR-FTIR measurements on glass substrates.

IR transmittance spectra were recorded using a Bruker Vertex 70 FTIR spectrometer at room temperature. The spectrometer is used with a ATR MIRacle (Pike) mono-reflection plate and a MCT detector (cooled with liquid nitrogen). The incidence angle is fixed at 45°. 1024 scans between 4000 and 600 cm$^{-1}$ with a resolution of 2 cm$^{-1}$ were recorded and averaged.

All the spectra were analyzed using OPUS software. The baselines were corrected using OPUS baseline correction. Transmittance bands are reported in reciprocal centimeters (cm$^{-1}$).

**X-ray Photoelectron Spectroscopy (XPS)**

The chemical composition of the coating on glass has been evaluated by XPS.

This measurement was performed under high vacuum (1.3 × 10$^{-9}$ mbar) with a Kratos Axis Ultra DLD spectrometer (monochromatic Al Kα (1486.6 eV) X-ray source) and a charge compensation system. A pass energy of 40 eV have been used for core levels spectra (high resolution, ± 0.5 eV). The X-ray beam is used to ionize on- surface atoms and a detector records the chemical shifts due to the ejected photoelectrons.

The data were analyzed with CasaXPS software. The peaks fitting was performed after subtracting a Shirley background. Chemical shifts are reported in electron volt (eV). All spectra were calibrated using the Si $2p_{3/2}$ band fixed at 103.5 eV.

# 3. Results and discussion

## 3.1. Surface topography

The surface topographies of DMDMOS-based coatings have been observed by AFM and ellipsometry mapping.

At small scale (5 µm x 5 µm), AFM measurements have been carried out on four coatings made under RH = 25 or 85 % and rinsed with $H_2O$+iPrOH or $H_2O$+iPrOH+toluene. The detailed results and experimental data can be found in *Supporting Information* (Figure SI.B.1). A roughness $R_q$ of 0.14 ± 0.01 nm is measured on the RH = 85 %, $H_2O$+iPrOH+toluene rinsed sample (smooth and homogeneous at this scale), which is in agreement with McCarthy *et al.*'s measurement, on a similar coating ($R_q$ = 0.10 ± 0.01 nm for RH = 65 %, same rinsing process) [2].

At larger scale (700 µm x 700 µm), ellipsometric mapping measurements evidence a spectacular humidity-dependent surface structuration (Figure 2). After a $H_2O$+iPrOH rinsing of a coating made under low humidity (RH = 25 %, Figure 2.a), islands of about 100 µm in diameter and large thickness (over 100 nm) are observed. The islands could not be imaged by ellipsometry because of the large slopes at their edges. Their summits are marked by red stars. By AFM (*Supporting Information*, Figure SI.B.2), the elevation of these islands from the coating background was measured in between 120 – 140 nm. As the ellipsometry mapping indicates a coating background thickness in between 10 – 50 nm (RH = 25 %, Figure 2.a), the islands are likely to be 130 to 190 nm thick from the silica layer of the substrate (background + over-thickness). For the coating made under high humidity (same rinsing process, RH = 85 %, Figure 2.a), islands of diameters of around 10 to 50 µm and of thicknesses in between 10 – 15 nm are observed by ellipsometry mapping, with a 3 – 7 nm thick background.

After a $H_2O$+iPrOH+toluene rinsing of the coating made under RH = 25 % (Figure 2.b), polydisperse round islands of thicknesses in between 10 – 40 nm and of diameters of around 1 to 30 µm are evidenced. A slight inhomogeneity is observed between the top left and the bottom right parts of the RH = 25 % image. This heterogeneity might be due to a heterogeneous rinsing of on-surface adsorbed molecules (see below). The thickness of the background decreases from 10 – 50 nm after a $H_2O$+iPrOH rinsing to 8 – 15 nm after a $H_2O$+iPrOH+toluene rinsing. For the coating made under RH = 85 % (same rinsing process, Figure 2.b), thinner (2 – 3 nm) and larger diameter (10 – 50 µm) islands than at low RH are observed. The thickness of the islands is also more homogeneous for the coating made under the highest humidity (for the same rinsing process, *i.e.* $H_2O$+iPrOH+toluene). In this latter case, the diameters of the islands are similar to the ones observed after the $H_2O$+iPrOH rinsing. Additionally, for the coating made under RH = 85 %, the thickness of the background decreases from 3 – 7 nm after a $H_2O$+iPrOH rinsing to 1 – 2 nm after a $H_2O$+iPrOH+toluene rinsing.

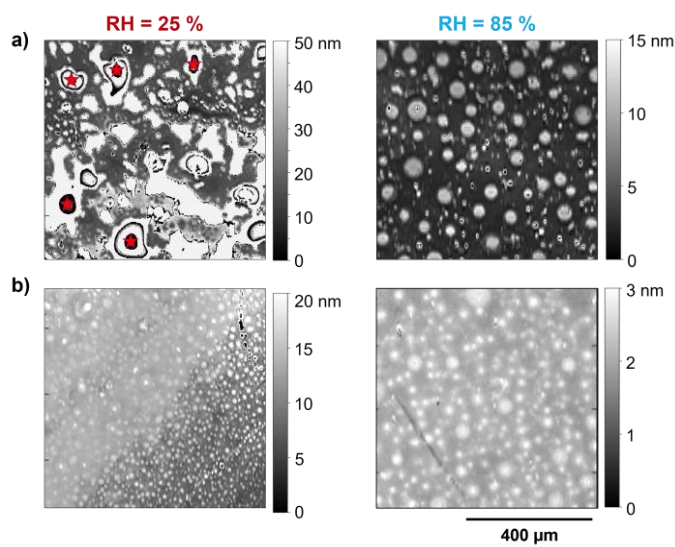

*Figure 2.* Ellipsometry mapping images of two DMDMOS-based coatings made under RH = 25 or 85 %, after **a)** a $H_2O$+iPrOH rinsing or **b)** a $H_2O$+iPrOH+toluene rinsing. The lateral scale is at the bottom right (identical for all images). The vertical scale (thickness) is different for each image.

The coating solution is composed of DMDMOS, $H_2SO_4$ and iPrOH. When deposited onto a substrate and placed in a humidity-controlled chamber, iPrOH evaporates, inducing a concentration of on-surface acidity. Some water molecules also transfer from the environment humidity to the coating layer. In these conditions (high acidity with water molecules), the polymerization of siloxanes (DMDMOS) is highly catalyzed [35,44]. Oligomers and polymers are promptly formed at the surface of the substrates. Thus, we herein assume that the coating layers are made of oligomers and polymers (PDMS-like molecules).

Moreover, the comparison of the ellipsometric mapping of the two rinsing processes shows that the thickness of both RH = 25 or 85 % coatings decreases after the use of toluene as a rinsing solvent. Toluene is a good solvent to solubilize PDMS-like molecules, even long polymer ones, unlike iPrOH [45]. Thus, as a large part of the coating is removed from the surface (on the background and on islands) after the use of toluene on both samples, we suggest that these rinsed molecules are likely to be weakly adsorbed PDMS, in particular long polymers which cannot be solubilized in iPrOH [4]. It can be highlighted that the amount of weakly adsorbed polymers in the coating made under low humidity (RH = 25 %) is significantly higher than in the coating made under high humidity. The outer layer of the coating made under RH = 25 % is less rinsed by iPrOH than the RH = 85 % one. After a $H_2O$+iPrOH rinsing, coatings are herein composed of weakly adsorbed polymers on top of a polymer background [4,46]. This latter structure (background and remaining islands, Figure 2.b), insensitive to rinsing, but whose thickness is highly sensitive to humidity, is consequently likely to be composed of grafted or irreversibly adsorbed polymers.

Finally, by comparing samples in Figure 2, we conclude that increasing the relative humidity used during the condensation step leads to thinner DMDMOS-based coatings, whatever the rinsing process used. Our results are in agreement with Wang *et al.*'s as they have measured, by ellipsometry, the thickness of coatings formulated from DMDMOS under the same conditions as ours ($H_2O$+iPrOH+toluene rinsing) and they have shown that this parameter decreases by increasing the condensation relative humidity [2]. However, we need to keep in mind that a classical ellipsometry measurement (at one location on the sample) only allows to obtain an average value of the thickness, as we have shown that the surface is not homogeneous.

All these results provide two key information regarding the molecular organization of DMDMOS-based coatings: 1) the grafted or strongly adsorbed molecules are longer when the coating is made under lower RH (thicker coatings); 2) the weakly adsorbed molecules are longer when the coating is made under lower RH (lower solubility in iPrOH).

A humidity-dependent topography on silicon-based coatings has already been evidenced by Artus *et al.* on coatings made from trichloroethylsilane with a Chemical Vapor phase Deposition (CVD) process [47]. By Scanning Electron Microscopy (SEM), they observed polysiloxane filaments at low RH (20 %) and wide polysiloxane rings at high RH (80 – 95 %) on the surface of their samples. However, our coatings are deposited in liquid phase, on the contrary of Artus *et al.*'s ones, deposited in gas phase. Our system is also composed of a bi-functional siloxane (dimethoxy) whereas theirs uses a tri-functional one (trichloro). The formation mechanisms of our coatings should then be different than theirs. The formation mechanism of silicon-based coatings in liquid phase have been briefly described by Wang *et al.* [2] but this mechanism is not yet well understood and requires further study. Thus, trough the study of the wetting and the chemical properties of our coatings, we will aim to draw the coating formation mechanism of these coatings.

## 3.2. Wetting and sliding properties

The wetting and sliding properties have been evaluated with a goniometer on the samples previously observed by ellipsometry mapping (RH = 25 or 85 %, $H_2O$+iPrOH or $H_2O$+iPrOH+toluene rinsing). The values obtained for the measurement of the contact angle CA, the hysteresis CAH and the sliding angle SA on these samples for droplets of water ($H_2O$), DMSO and tetradecane are presented in Figure 3.

It first appears that CAH and SA obtained on $H_2O$+iPrOH rinsed samples are smaller than those on $H_2O$+iPrOH+toluene rinsed samples for all the tested liquids and for both RH. However, for RH = 25 % $H_2O$+iPrOH rinsing (Figure 3.a), CAH and SA could not be measured for tetradecane droplets as this liquid totally wets the surface with time. The $H_2O$+iPrOH rinsing process of DMDMOS-based coatings leads to the smallest sliding angles and thus improves the slippery behavior of these coatings. It is remarkable that such low values of hysteresis and sliding angles of about 5° are obtained in these conditions, despite the evidenced substrate nanostructuration. Regarding the value of CA, it remains stable between both rinsing processes for $H_2O$ droplets, while it increases slightly from a $H_2O$+iPrOH rinsing to a $H_2O$+iPrOH+toluene rinsing for DMSO droplets, and more significantly for tetradecane droplets. It has previously been shown that after a $H_2O$+iPrOH rinsing, coatings are composed of two parts: an internal layer of grafted polymers covered by an external layer of weakly adsorbed polymers. DMSO and even more tetradecane have the ability to solubilize PDMS-like molecules. Thus, the deposition of droplets of these liquids on $H_2O$+iPrOH rinsed coatings might lead to the addition of some weakly adsorbed molecules of the coating to the liquid droplets. This addition of PDMS ($\gamma_{SV,PDMS}$ = 20 – 24 mN·m$^{-1}$ [15,16]) lowers the surface tension of the liquid from the droplet, leading to smaller CA values according to Young's equation [48]: $\cos(\text{CA}) = \frac{\gamma_{SV} - \gamma_{SL}}{\gamma_{LV}}$, with $\gamma_{SV}$, $\gamma_{SL}$ and $\gamma_{LV}$, the surface tensions at the solid/gas interface, at the solid/liquid interface and at the liquid/gas interface, respectively. The CA values from $H_2O$+iPrOH rinsed coatings to $H_2O$+iPrOH+toluene rinsed ones increases as the droplet is enriched by PDMS ($\gamma_{LV}$ decreases).

Moreover, by comparing both RH = 25 % and RH = 85 % samples for a same rinsing process, it is clear that using a high relative humidity during the condensation step leads to a more slippery coating towards water, DMSO and tetradecane droplets (lower SA and CAH). We hence confirm the results previously reported by Wang *et al.* [2], *i.e.* that increasing RH decreases SA. Our experiments extend theirs by reaching a higher humidity (they only studied RH to 65 %) and evaluating the SA for two other liquids than water. Combining these two findings, on a DMDMOS-based coating made under RH = 85 % and with a $H_2O$+iPrOH rinsing, CA = 105 ± 1°, CAH = 4 ± 2° and SA = 5 ± 2° for droplets of water; CA = 72 ± 1°, CAH = 7 ± 2° and SA = 8 ± 1° for droplets of DMSO; CA = 17 ± 1°, CAH = 4 ± 1° and SA = 1 ± 1° for droplets of tetradecane.

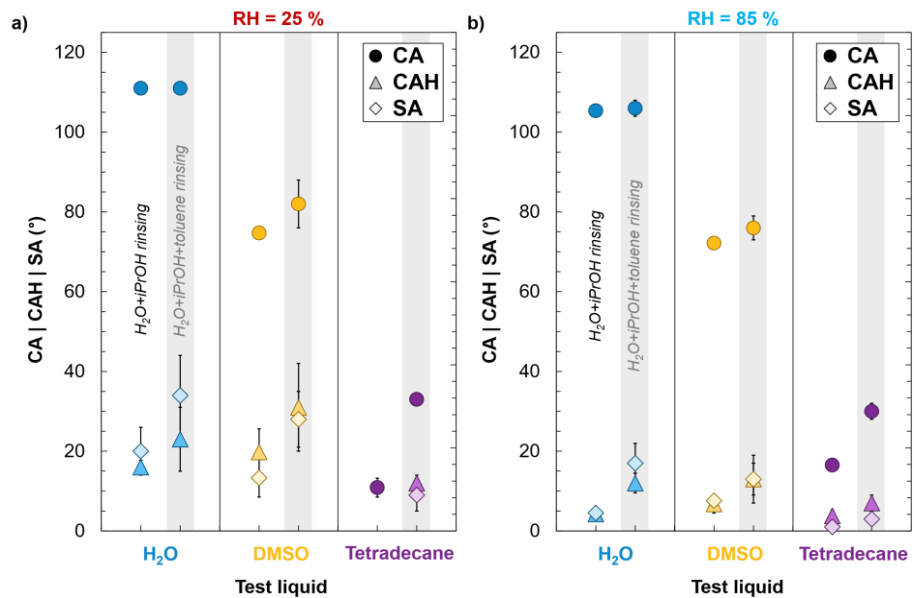

*Figure 3.* Contact angle (CA, circular symbols), hysteresis (CAH, triangular symbols) and sliding angle (SA, diamond symbols) for water ($H_2O$), DMSO and tetradecane droplets deposited on DMDMOS-based coatings made at **a)** RH = 25 % or **b)** 85 %, and after a $H_2O$+iPrOH rinsing (transparent background) or a $H_2O$+iPrOH+toluene rinsing (grey background). If error bars are not visible for some measurements, it means the standard deviation is small and error bars are smaller than the size of the symbols used. For RH = 25 % $H_2O$+iPrOH rinsing, CAH and SA could not be measured for tetradecane droplets as this liquid totally wets the surface with time.

The evolution of CA, CAH and SA on $H_2O$+iPrOH+toluene rinsed samples as a function of the condensation RH (8, 25, 30, 50, 65 and 85 %) has been tested in greater detail in Figure 4, validating that increasing RH decreases the values of CAH and SA.

Moreover, it has been found that increasing the relative humidity also slightly impacted CA values. A way to understand why the water contact angle varies with humidity is related to the overall thickness decrease with humidity, shown previously with ellipsometry mapping images. The thinner the coating, the stronger the interaction of water with the silica layer (and hydroxyls groups) on the substrate surface. This link between the coating thickness and CA have been emphasized by Bain *et al.* [49]. However, their study also indicated that CA stabilizes at a maximum thickness of 1.5 nm. In our study, the coating layers are thicker (up to 190 nm) showing that other effects probably contribute to the link between the layer thickness and the variation of CA. Since we previously evidenced a humidity-dependent topography, a relationship between wettability and topography, such as a Wenzel effect [50] could be suspected. Indeed, the latter effect means that an increase in roughness of a hydrophobic substrate leads to an increase of the contact angle. However, the contact angle obtained for water droplets on a flat PDMS surface being around 114°, in our samples, the roughness due to the nanostructuration would have induce an increase of this CA value which is contrary to our measurements (CA between 103° and 111°). Also, a Wenzel effect due to roughness changes with humidity should lead to identical variations for water, DMSO and tetradecane CA values with increasing humidity, which is contrary to the results shown in Figure 4. The decrease in the CA value with increasing RH is more significant for water droplets than for the two other tested liquids (slope of the linear regression = -0.07 versus -0.02 and 0 for DMSO and tetradecane drops, respectively). Therefore, we suggest that this effect could be related to the presence of a greater number of hydroxyl groups on the surface of the sample made under high humidity, thus making the coating slightly more hydrophilic [51]. This hypothesis could explain the decrease of CA with the increase of humidity as well as the difference of behavior between the liquids since the behavior of a polar liquid ($H_2O$) will be more impacted by the presence of hydroxyl groups on the surface, than a non-polar liquid (*i.e.* tetradecane).

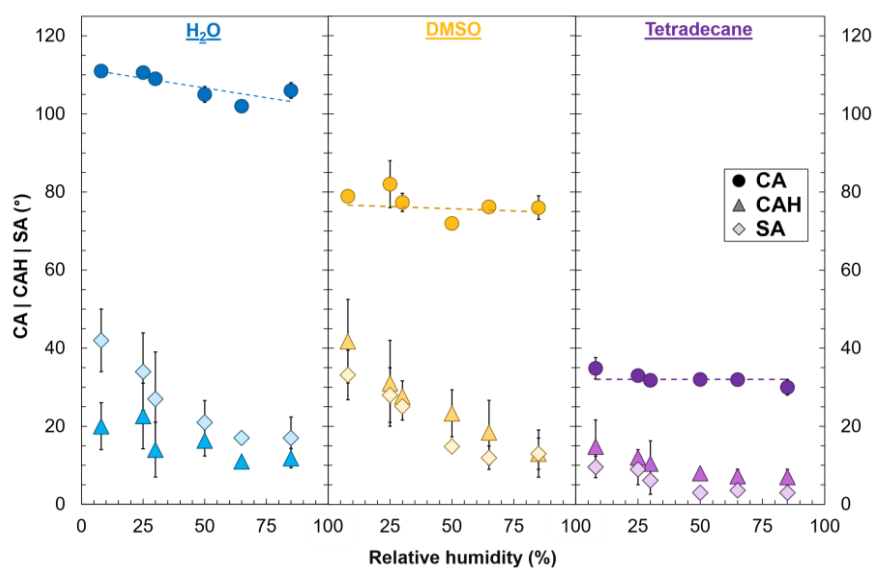

*Figure 4.* Contact angle CA (circular symbols), hysteresis CAH (triangular symbols) and sliding angle SA (diamond symbols) for water ($H_2O$), DMSO and tetradecane droplets as a function of the relative humidity imposed during the condensation step of DMDMOS coating process on glass slides (after $H_2O$+iPrOH+toluene rinsing). Dashed lines represent trends in CA variation (linear fit). If error bars are not visible for some measurements, it means the standard deviation is small and error bars are smaller than the size of the symbols used.

## 3.3. Chemical composition and molecular organization

The presence of hydroxyl groups inside or on the surface of the coatings has been evaluated through silylation experiments with HMDS, aiming at transforming hydroxyl groups in trimethylsiloxyl ones as illustrated in Figure 5.

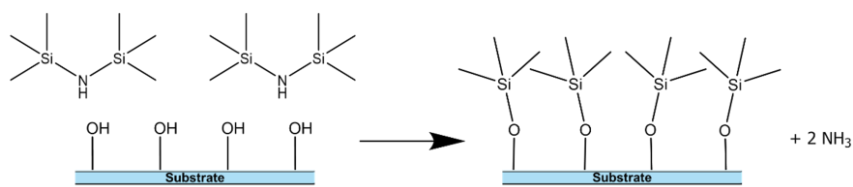

*Figure 5.* HMDS silylation mechanism of a substrate with hydroxyl groups (at room temperature).

Figure 6 shows the evolution of the contact angle CA before ("Initial") and after a silylation treatment of both coated samples RH = 25 or 85 %, H$_2$O+iPrOH+toluene rinsing, with three different processes. No significant effect of HMDS is observed on the CA value for both coatings and all silylation processes, showing there are very few hydroxyl groups available in these coatings or on surface of the substrate. Had such groups been found, the substitution of hydroxyl groups by silyl groups would have lowered the surface energy $\gamma_{SV}$ of the sample. Thus, an increase in the contact angle value CA should have been recorded, according to Young's equation [48].

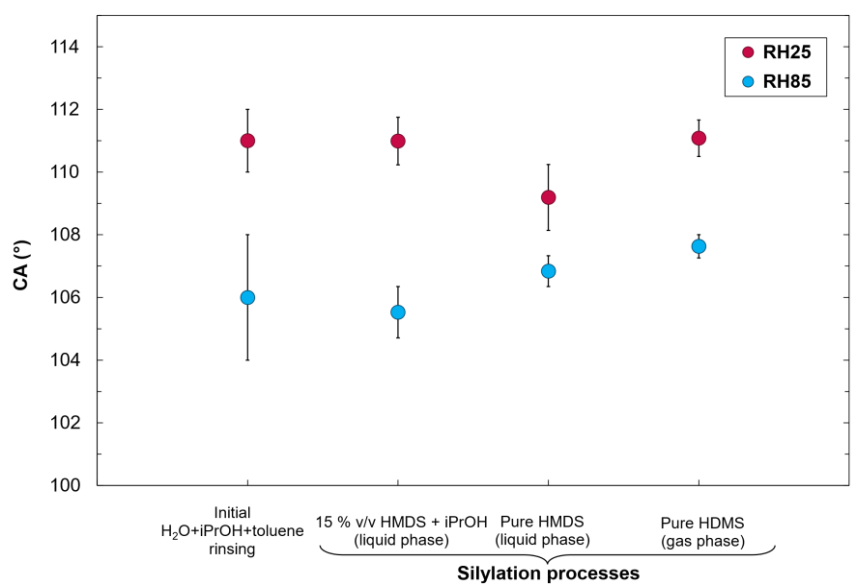

*Figure 6.* Contact angle CA for two DMDMOS-based coatings made under RH = 25 or 85 %, H$_2$O+iPrOH+toluene rinsing, before (initial) or after silylation by three processes: diluted in liquid phase (15 % v/v HDMS + iPrOH), pure in liquid phase or pure in gas phase.

ATR-FTIR and XPS measurements have also been performed on these samples (*Supporting Information*, Figure SI.C.1 and Figure SI.D.1). The presence of hydroxyl groups in these coatings could not be evidenced by ATR-FTIR (large band between 3200 – 3400 cm$^{-1}$ for bonded OH or a thin one at 3750 cm$^{-1}$ for free OH [52]). These experiments, though, confirm that PDMS-like molecules have been deposited on the substrates. However, no salient different feature between the different coatings at various humidities could be evidenced from a chemical composition's point of view by XPS and IR techniques (see *Supporting Information*). We therefore conclude that, for these DMDMOS systems, the chemical composition of the coating and the molecular organization vary little with the process.

As we also do not detect any methoxyl O–CH$_3$ functions on these coatings (ATR-FTIR signal supposedly located at σ = 820 cm$^{-1}$ [53] and XPS band at $E_L$ = 286,5 eV [54], Figure 7), we conclude that the coatings are not composed of PDMS brushes terminated with methoxyl groups as previously stated by Wang *et al.* [2]. In their previous work, they suggested that the DMDMOS-based coatings were formed by a grafting-from mechanism of hydrolyzed DMDMOS molecules. These polymer brushes are supposedly methoxyl-ended due to mono-hydrolyzed DMDMOS molecules. However, no methoxyl O–CH$_3$ functions are detected in these coatings. We have also evidenced that the contact angle value measured on the H$_2$O+iPrOH+toluene rinsed samples is stable after one year of storage at room temperature (CA = 111 ± 2° and 106 ± 2° for coating made under RH = 25 or 85 %, respectively). The presence of methoxyl-terminated brushes seems then controversial, as these molecules would have been easily hydrolyzed during this time lapse at ambient humidity, leading to a decrease in the contact angle value.

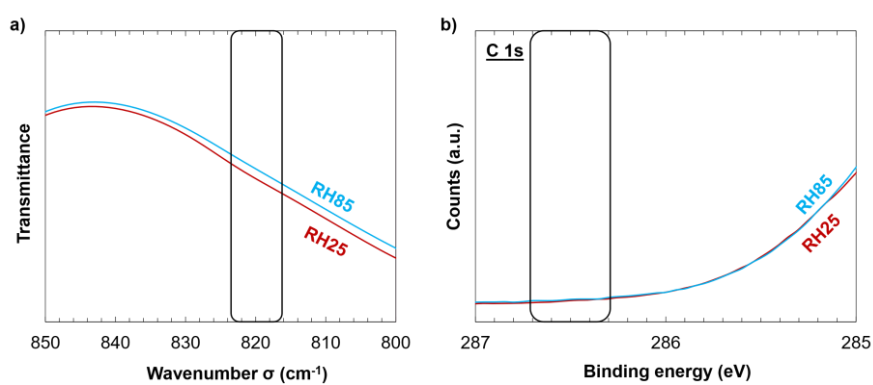

*Figure 7.* **a)** ATR-FTIR spectra between 800 and 850 cm$^{-1}$ and **b)** C 1s XPS spectra between 285 and 287 eV of DMDMOS-based coatings made under RH = 25 % (red curve, RH25) or RH = 85 % (blue curve, RH85), with a H$_2$O+iPrOH+toluene rinsing. The black boxes highlight the zones where the methoxyl O–CH$_3$ signal is expected. In figure a) the spectra were obtained after correction of the baseline with OPUS correction tool. In figure b), the spectra are overlapping.

### 3.4. Coating formation mechanisms

We previously suggested that polymers could be rapidly formed on the surface of the samples during the condensation step due to the thin film deposition and the acidity of the solution. The formation of polymers along with the evaporation of iPrOH during condensation can lead to other polymer conformations.

After the deposition of the coating solution on the glass slide, various on-surface reactions can occur depending on the relative humidity (Figure 8.a, 7.c, 7.e). As water is a catalyst for siloxane polymerization [35,44], the higher the relative humidity used during the condensation step, the faster the polymerization of siloxanes. In their study, Artus *et al.* suggest that, in a gas phase process, water droplets condense on the substrates and act as reaction vessels. Silanes added to the reactive medium are then attracted and dissolved in these water droplets. Polymerization and grafting mechanisms onto the substrate finally occur to form surface structures. In our case, the coating solution deposited on surface of the substrate is composed of iPrOH, acid and DMDMOS-based molecules. Water from relative humidity can not condense on the substrate but is added to the deposited coating solution.

At low RH (RH = 25 %, Figure 8.a), few water molecules enter the deposited coating solution. The water content is low such as polymers need to reach a threshold in molecular weight to induce a phase separation. Moreover, in this process configuration, the polymerization kinetics is slow (low water content). Two competitive mechanisms can be evidenced: first, the formation and the phase separation of polymers from the iPrOH solution, and, secondly, their grafting to the hydroxyl groups of the silica on the substrate.

At high RH (RH = 85 %, Figure 8.a), as water molecules enrich the main phase, two competitive mechanisms can occur. On one side, the polymerization is highly catalyzed by water so long polymers are rapidly formed; on the other side, the solubility of PDMS-like molecules decreases with the increasing water ratio to the main phase so that a phase separation is observed between polymers and the solution (iPrOH, water, acid, oligomers) at a lower threshold than for the RH = 25 % case. Therefore, at this stage, the main phase of the coating at RH = 25 % is supposedly composed of longer oligomers than the main phase at RH = 85 %, the longest one being rapidly phase-separated from the main phase.

Through the condensation step, the solvents of the main phase (iPrOH and water) evaporate, inducing an increase in oligomer concentration and acidity. This evaporation is faster at low humidity due to the dryness of the environment. The high concentration of the continuous phase on the substrate favors the adsorption of oligomers as Guiselin brushes [32] (loops, Figure 8.b) as demonstrated by Cohen-Addad *et al.* [33] and Auvray *et al.* [34]. In our study, these oligomers are hydrogen bonded between the hydroxyls of the silica and the oxygen atoms of the polymers, and, for most of them, covalently bonded at the chain ends (Figure 8.b), explaining why the silylation experiment, previously detailed, was not efficient. Moreover, Auvray *et al.* have shown that when Guiselin brushes are adsorbed on a substrate from a concentrated solution (as in our study), the grafting density is high so most of the hydroxyl groups of the silica surface are involved in chemical bonds (hydrogen or covalent ones) [34]. As oligomers in the main phase at RH = 25 % are longer than those at RH = 85 %, the background formed is thicker at RH = 25 % than at RH = 85 % [32,33], as shown previously with the ellipsometric mapping observations (Figure 2).

The evaporation of the solvents also induces a deposition of the phase-separated polymer domains on the background already adsorbed in Guiselin brushes (Figure 8.c). The polymers of the domains may coalesce with the latter brushes, suppressing the Guiselin brush/main phase interface at the deposition zones. These polymer domains thus can limit the hydrolysis of these Guiselin brushes by inhibiting the interactions with the water from the main phase. Outside of these areas, water can hydrolyze the loops (Figure 8.c-e), resulting in a decrease in coating thickness, and allow the formation of the characteristic structures previously observed. Hydrolysis of the Si–O–Si bonds of the loops splits them into two OH-terminated brushes (Figure 8.d). This hydrolysis is more significant and homogeneous at high humidity than at low humidity, explaining the differences in surface structures. The split loops can immediately fold back onto the surface to covalently bond with the silica (if free hydroxyls remain) or associate with free oligomers in the main phase, lengthening the chain of the grafted polymer (Figure 8.f). These new chains can then fold back onto the surface or lengthen again.

After the evaporation of all the solvents from the main phase, an equilibrium state is reached, and a gel is formed on the substrates. For the coating made under RH = 25 %, this gel is composed of long oligomers and polymers. For the one made under RH = 85 %, the gel is composed of shorter oligomers and polymers. Both coatings are then composed of two layers: an inner layer of Guiselin brush end-grafted polymers and an outer layer of weakly adsorbed polymers (Figure 8.e).

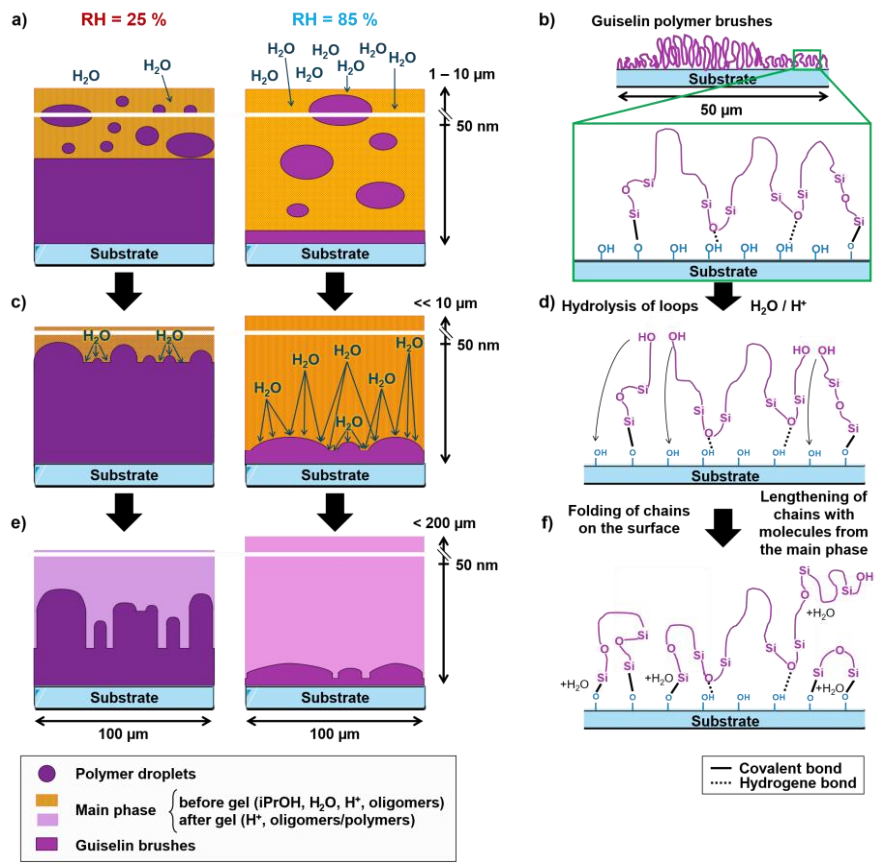

*Figure 8.* Proposed reaction schematics for the formation of DMDMOS-based coatings as a function of RH. Three steps are described as follows: **a)** water intrusion in the main phase due to relative humidity (represented by arrows), inducing an on-surface phase separation between polymers and the main phase. Some of the formed polymers are directly grafted to the substrate in a Guiselin brush conformation; **c)** deposit of the polymer droplets on the grafted polymers and hydrolysis of these grafted polymers outside the droplets (represented by arrows from water molecules to the grafted polymer layer); **e)** gel of the main phase (solvent evaporation) and formation of the islands. The vertical scale indicates the range of theoretical values of main phase thickness for a dip-coating withdrawal rate of 0.1 mm·s$^{-1}$. This thickness evolves with solvent evaporation. The white horizontal line represents a break in the scale. Figures **b)**, **d)** and **f)** show reactional schematic of scission of PDMS loops by hydrolysis, and potential behaviors of the chains formed after scission of the loops: folding on the surface to form a new smaller loop (represented by arrows) or lengthening by condensation of a molecule from the main phase. These reactions occur during the phase represented in figure c).

After condensation, several rinsing steps are performed. Water is first used to stop the reaction by quenching the acidity from the surface. The following iPrOH rinsing (Figure 9.a) solubilizes the adsorbed short chains, removing a significant portion of the adsorbed layer at high humidity. For the coating made at low humidity, the chains of this outer layer being longer, they are more entangled and more difficult to desorb. For both coatings made at RH = 25 or 85 %, this rinsing step leaves a layer of oligomers of varying lengths on the surface. This weakly adsorbed layer covers the inner (strongly adsorbed) layer. The toluene rinsing (Figure 9.b-c) removes the remaining adsorbed molecules in the outer layer by solubilization. However, if free polymer chains are entangled in the Guiselin brushes, they may remain in the coating after this rinsing step, as shown by Teisala *et al.* [4].

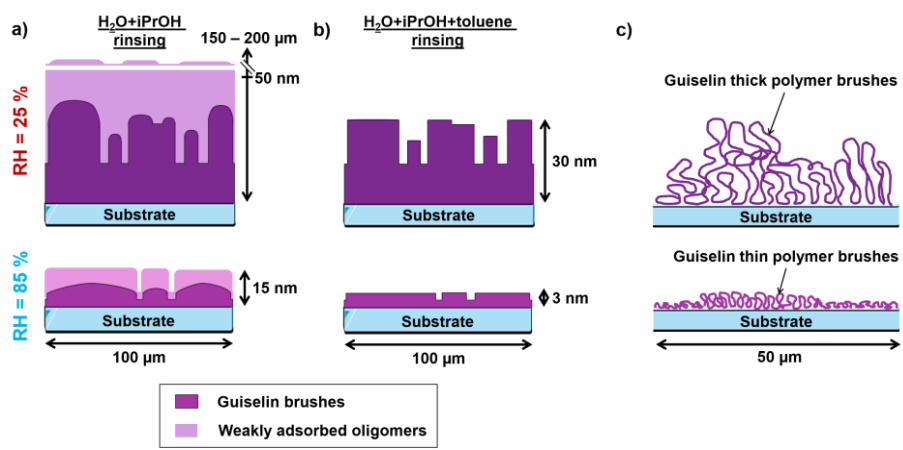

*Figure 9.* Schematic of the surface topography of DMDMOS-based coatings made under RH = 25 or 85 %, **a)** after a $H_2O$+iPrOH rinsing, and **b)** after a $H_2O$+iPrOH+toluene rinsing. **c)** Schematics showing molecular Guiselin brushes organization of polymers on the substrates after a $H_2O$+iPrOH+toluene rinsing and as a function of RH. The vertical scales have been set according to the ellipsometry mapping results.

## 3.5. Link between chemical organization, surface topography and sliding

Even though we did not detect any hydroxyl groups in the coatings, some might remain in a very low quantity, which makes it hard to be identified by ATR-FTIR. These groups might be located on the background of the coatings (outside of the islands) or still on the silica layer. HMDS molecules might hardly access these functional groups, due to possible high brush density. Considering a Guiselin brush organization with shorter polymers at high humidity than at low humidity, we can assume that the chain tail to loop ratio of Guiselin brushes and their grafting density will be higher at higher humidity [55]. A high grafting density limits the refolding of chain tails on the substrate and the formation of covalent bonds. More hydroxyl groups located on the tails of the polymers could therefore be able to be in contact with the liquid of the droplet when deposited on the coating made at higher humidity. These interactions would explain the slight differences in contact angle observed previously between these two coatings. However, this hypothesis still needs to be precisely verified as the surface structuration could restrain interactions between the droplet and the background of the substrate.

This surface nanostructuration is also striking since it induces a low hysteresis despite heterogeneity when the coating is made at high relative humidity (for water droplets, CAH = 12 ± 2° after a $H_2O$+iPrOH+toluene rinsing). Indeed, wetting properties, and in particular hysteresis, are intimately linked to the surface heterogeneities. Geometrical heterogeneities on the coating surface can explain high hysteresis or sliding angles values as they act as pinning points for droplets [56]. However, in this study, samples made under both low and high humidities showed structured surfaces with specific patterns related to different hysteresis values. We can then hypothesize that the main parameters influencing the hysteresis seems to be the size and thickness of these surface heterogeneities. Precisely, wide and thin structures lead to smaller hysteresis and more slippery samples than small and thick islands. As composed of Guiselin brushes, these structures are not rigid as they might be in physically nanostructured surfaces (by laser ablation [39,57], etching [36]): here, the flexibility of polymer brushes induces an increased mobility of the chains that prevents the interface pinning, so the surface structuration is not an issue to sliding.

The sliding differences are related to differences in chain mobilities, which are effective to create more or less liquid-like islands. For long Guiselin brushes (lower RH), an entanglement of the polymers can be observed [58,59]. This entanglement could inhibit the mobility of the brushes and the flexibility of the Si–O–Si bonds, explaining why the coating made at RH = 25 % is less slippery than the one made at RH = 85 % (with $H_2O$+iPrOH+toluene rinsing). For $H_2O$+iPrOH rinsed samples, the weakly adsorbed polymers above the Guiselin brushes (outer layer) act like a liquid layer and moves freely on the surface, allowing a better sliding of the droplets. This organization can be related to a combination of "pseudo-brushes" (Guiselin brushes) and of a "bimodal" model (adsorbed chains on-top of grafted brushes) as presented by Casoli *et al.* [46].

# 4. Conclusions

Harmless and eco-friendly silicon-based omnirepellent coatings based on silicones have been investigated in this work. This recent approach has first been pioneered by McCarthy's group, creating coatings from one-side grafted silicon brushes [2]. Having a full understanding of these systems is highly recommended if we aim to optimize the coating process and transpose this study to other siloxane molecules or other substrates. However, no clear evidence of the molecular organization in these coatings was given so far.

Our purpose was to identify the polymer on-surface conformation and to demonstrate the formation mechanism of this coating. A multiscale approach has been conducted from the study of the macroscopic properties (sliding, contact angles) and of the microscopic properties (topography) to the characterization of the molecular composition. A spectacular humidity-dependent surface structuration is discovered that we link to a phase separation between polymers and the main phase of the coating (acid, solvent, oligomers) at the surface of the substrate. With solvent evaporation, the oligomers concentration increases in the main phase, favoring their adsorption as Guiselin brushes [32] (loops) on the substrate surface. A complex mechanism of hydrolysis of the loops and refolding on the surface was detailed to explain the formation of the surface structures. The coating made under low humidity is the thickest [32,33]. We finally demonstrated that the increase of the sliding behavior with increasing relative humidity is related to differences in chain lengths and chain mobilities: the smaller the loops, the lower the risk of entangling so the higher their mobility and the lower the sliding angle. By comparison with recent literature, our contribution can extend Artus *et al*'s work on polysiloxane-based gas phase deposited coatings [47], as we propose a liquid phase polysiloxane-based coating formation mechanism. Moreover, we correct and complement results by McCarthy *et al*. by elucidating the molecular structure and the large-scale surface structuration [2]. It is remarkable that a very small hysteresis (below 5°) is reached despite a clear nanostructuration.

If weakly adsorbed polymers remain on the surface, above the Guiselin brushes, they would act like a liquid, increasing the sliding of the droplets up to a sliding angle of less than 5° for water, DMSO and tetradecane droplets. This weakly adsorbed polymer layer seems to be the key parameter to increase the slippery behavior of this coating. However, preliminary results have shown that this layer is quite unstable as it can be removed by using a good solvent for PDMS (toluene, tetradecane), a long immersion in an intermediate solvent (isopropanol) or even by the rolling/sliding of many drops of a bad solvent (water). Cross-linking processes are now under scrutiny to gently stabilize this layer and preserve the sliding properties, without interfering with the Guiselin brushes. For industrial interest (cost, accessibility), the coating process used with DMDMOS-based coatings will be transposed on formulas composed of cyclic siloxanes. Early results are promising, but the process needs to be carefully adjusted as the hydrolysis and polymerization kinetics of these cyclic siloxanes is different of DMDMOS ones. These systems still need more study.

To conclude, the better understanding of the formation mechanism of the siloxane-based coatings can pave the way to the design of omnirepellent coatings, more respectful of the environment, economical, durable and harmless.

## Acknowledgements

We want to thank the *Commissariat à l'énergie atomique et aux énergies alternatives* (CEA) of Saclay (France) for funding this project. We also thank Ludovic Tortech (*Laboratoire d'Innovation en Chimie des Surfaces et Nanosciences* (LICSEN) – CEA Saclay) for his help on the AFM experiments. We finally thank Alban CHARLIER for the preliminary results obtained during his internship.

# Supporting Information

## A – Choice of the hysteresis measurement method

Four hysteresis (CAH) calculation methods have been compared on DMDMOS-based coatings made under RH = 25 or 85 % ($H_2O$+iPrOH rinsing). For the first needle method, a 2 µL droplet has been deposited on both samples, inflated at 0.2 µL·s$^{-1}$ to 22 µL, stabilized for 2 min and then deflated at 0.2 µL·s$^{-1}$ to 2 µL before a 2 min stabilization phase. The first hysteresis calculation method with the needle experiment has been presented by Wong *et al.* [60]: CAH is measured by subtracting the receding angle θ$_r$ obtained at the end of the deflection of the droplet, to the advancing angle θ$_a$ obtained at the end of the inflation of the droplet. The second hysteresis calculation method with the needle experiment has been presented by Barrio-Zhang *et al.* [61]: CAH is measured by subtracting the receding angle θ$_r$ obtained after the deflection and the stabilization steps, to the advancing angle θ$_a$ obtained after the inflation and the stabilization steps. The third hysteresis calculation method with the needle experiment consist of subtracting the receding angle θ$_r$ obtained when the contact line between the sample and the droplet begins to move during deflection, to the advancing angle θ$_a$ obtained when the contact line between the sample and the droplet begins to move during inflation. This method will be identified as "contact line method". For the tilting plate method (fourth method), the process used is described in part 2.4 of the Materials and methods section.

The CAH values obtained by these calculation methods for droplets of water are presented in Table SI.A.1. These results are rather scattered in the determination of CAH depending of the calculation method used.

*Table SI.A.1.* Hysteresis CAH values obtained by four calculation methods for droplets of water deposited on DMDMOS-based coatings made under RH = 25 or 85 % ($H_2O$+iPrOH rinsing).

| Sample | CAH by needle method (inflation/deflation of droplet) (°) | | | CAH by tilting plate method (°) |
|---|---|---|---|---|
| | **Wong** [60] | **Barrio-Zhang** [61] | **Contact line method** | |
| **RH = 25 %** | 21 | 9 | 13 | 16 ± 2 |
| **RH = 85 %** | 14 | 7 | 5 | 4 ± 2 |

To determine the most accurate CAH calculation method for our study, a comparison with the theoretical sliding angle SA* provided by the Furmidge equation (A.1) has been calculated for each CAH value and compared to the experimental sliding values SA.

$$sin(SA) = \frac{\omega \gamma_{LV}}{mg}(cos\,\theta_r - cos\,\theta_a) \qquad \textbf{(A.1)}$$

with $\omega$, contact line diameter; $m$, mass of the droplet; $\gamma_{LV}$, surface tension of the droplet liquid (here, water so $\gamma_{LV,\,20\,°C}$ = 72,8 mN·m$^{-1}$ [62]); $g$, gravitational acceleration ($g$ = 9.8 m·s$^{-2}$).

However, to use the Furmidge equation, the contact line diameter $\omega$ needs to be determined. By using trigonometry (Figure SI.A.1), $\omega$ can be expressed as a function of the contact angle CA and the volume of the deposited droplet $V_g$ as follows:

$$\omega = 2 \sin CA \sqrt[3]{\frac{3V_g}{\pi(4-(1+\cos CA)^2(2-\cos CA))}} \qquad \textbf{(A.2)}$$

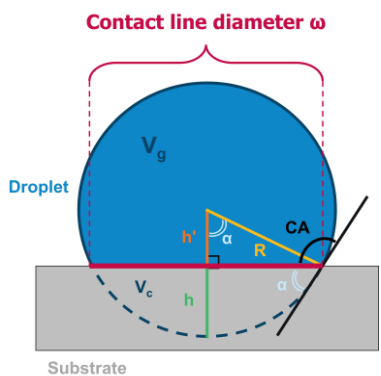

***Figure SI.A.1.*** Schematics of a droplet deposited on a substrate with the trigonometric parameters needed to determine the expression of the contact line diameter $\omega$.

The calculated theoretical sliding angles SA* are presented in Table SI.A.2. For both samples, the theoretical SA* values are closest to the experimental SA values obtained by the tilting plate method. Interestingly, these conclusions disagree with Krasovitski *et al.* [63] where they advise against using this tilting plate method to determine CAH! Indeed, their work shows that, for high contact angles (> 100°), measuring an advancing angle by analyzing the shape of a drop on an inclined surface does not correspond to the calculated limit value of the maximum angle that this drop can form. However, their study involves much larger sliding angles than ours and surface chemical heterogeneities, suggesting that their model is not applicable to our study.

***Table SI.A.2.*** Theoretical sliding angles SA* calculated from hysteresis CAH values obtained by four calculation methods for droplets of water deposited on DMDMOS-based coatings made under RH = 25 or 85 % ($H_2O$+iPrOH rinsing).

| Sample | SA* for CAH values obtained by needle method (inflation/deflation of droplet) (°) | | | SA* for CAH values obtained by tilting plate method (°) | Experimental SA (°) |
|---|---|---|---|---|---|
| | **Wong** [60] | **Barrio-Zhang** [61] | **Contact line method** | | |
| RH = 25 % | 27 | 11 | 16 | 20 | 20 ± 6 |
| RH = 85 % | 19 | 9 | 7 | 5 | 5 ± 2 |

## B – AFM measurement

The roughness ($R_q$) and surface topography at small (5 x 5 µm) or medium (100 x 100 µm) scales were evaluated through AFM measurements. The Nano-Observer AFM from CSInstruments was used on silicon wafer samples (bare and coated). The resonant (non- contact) mode was selected and PPP-NCL tips (resonance frequency = 190 kHz, force constant = 48 N·m$^{-1}$) from Nanosensors have been used. Images were processed and analyzed with Gwyddion software.

Figure SI.B.1 shows AFM 2D images of coatings made under RH = 25 or 85 % and after a $H_2O$+iPrOH rinsing or a $H_2O$+iPrOH+toluene rinsing (full scale, 5 x 5 µm). An identical

roughness $R_q$ of 0.10 ± 0.01 nm is measured for both samples made under RH = 25 or 85 % and H$_2$O+iPrOH rinsed (Figure SI.B.1.a). For these coatings, the surface is smooth and homogeneous at this scale. After a H$_2$O+iPrOH+toluene rinsing (Figure SI.B.1.b), $R_q$ slightly increases for RH = 85 % sample ($R_q$ = 0.14 ± 0.1 nm). For RH = 25 % sample, islands with a thickness of around 40 nm and a diameter of around 0.1 to 1 nm appear, increasing the value of $R_q$ up to 3.45 ± 1.24 nm. Nevertheless, these images cannot be directly correlated with the sliding properties, as they are not measured at the drop scale: for 20 µL drops, with high contact angles (105 – 111°), the contact line diameter is between 345 and 368 µm.

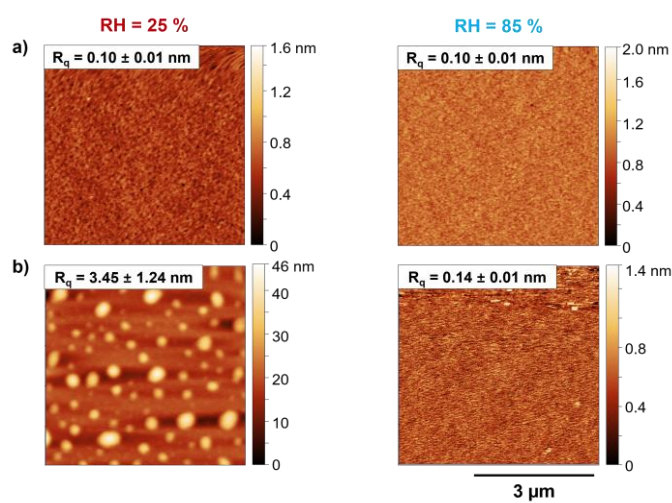

*Figure SI.B.1.* AFM 2D images of two DMDMOS-based coatings made under RH = 25 or 85 %, after **a)** a H$_2$O+iPrOH rinsing or **b)** a H$_2$O+iPrOH+toluene rinsing. The roughness values $R_q$ are listed at the top left of the images. The lateral scale is at the bottom right and is identical for all images. The vertical scale (thickness) is different for each image.

Figure SI.B.2 shows an AFM 3D image of islands with thicknesses of around 120 – 140 nm on a DMDMOS-based coating made under RH = 25 %, after a H$_2$O+iPrOH rinsing at a larger scale comparable to the droplet size.

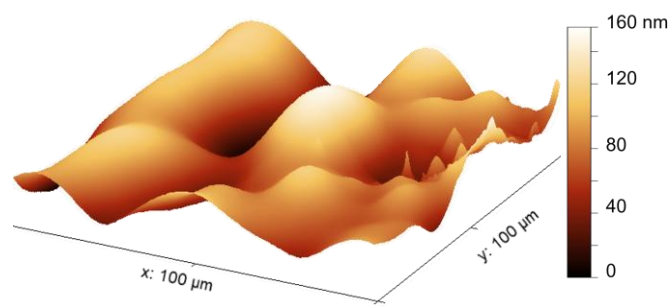

*Figure SI.B.2.* AFM 3D image of a DMDMOS-based coating made under RH = 25 %, after a H$_2$O+iPrOH rinsing.

## C – ATR-FTIR measurements

ATR-FTIR measurements were performed on the DMDMOS-based coatings (RH = 25 or 85 %, H$_2$O+iPrOH or H$_2$O+iPrOH+toluene rinsing) to assess the presence of PDMS-like molecules and of hydroxyl groups. Two characteristic peaks of the substrate (glass) have been detected at 760 cm$^{-1}$ and 914 cm$^{-1}$ (Figure SI.C.1.a, top curve, black), corresponding to the deformation of Si–O bond in SiO$_2$ and to the stretching of Si–O bond in Si–OH, respectively [35]. For the coated samples, peaks corresponding to the stretching of C–H bonds in Si–CH$_3$ located at 2963 cm$^{-1}$ (Figure SI.C.1.b) and to the deformation of Si–C bonds in Si–CH$_3$ located at 1260 cm$^{-1}$ (Figure SI.C.1.c) are detected. The deformation of Si–O bonds in Si–O–Si is also detected at 795 cm$^{-1}$, for all four coatings. All these signals proves the presence of PDMS molecules (O–Si(CH$_3$)$_2$) [64,65]. These ATR-FTIR spectra also highlight that the intensity of the peaks located at 2963 cm$^{-1}$ and 1260 cm$^{-1}$ is linked to the thickness of the coatings observed in ellipsometry mapping: the thicker the coating, the higher the intensity. The presence of hydroxyl groups in these coatings could not be evidenced here (large band between 3200 – 3400 cm$^{-1}$ for bonded OH or a thin one at 3750 cm$^{-1}$ for free OH [52]).

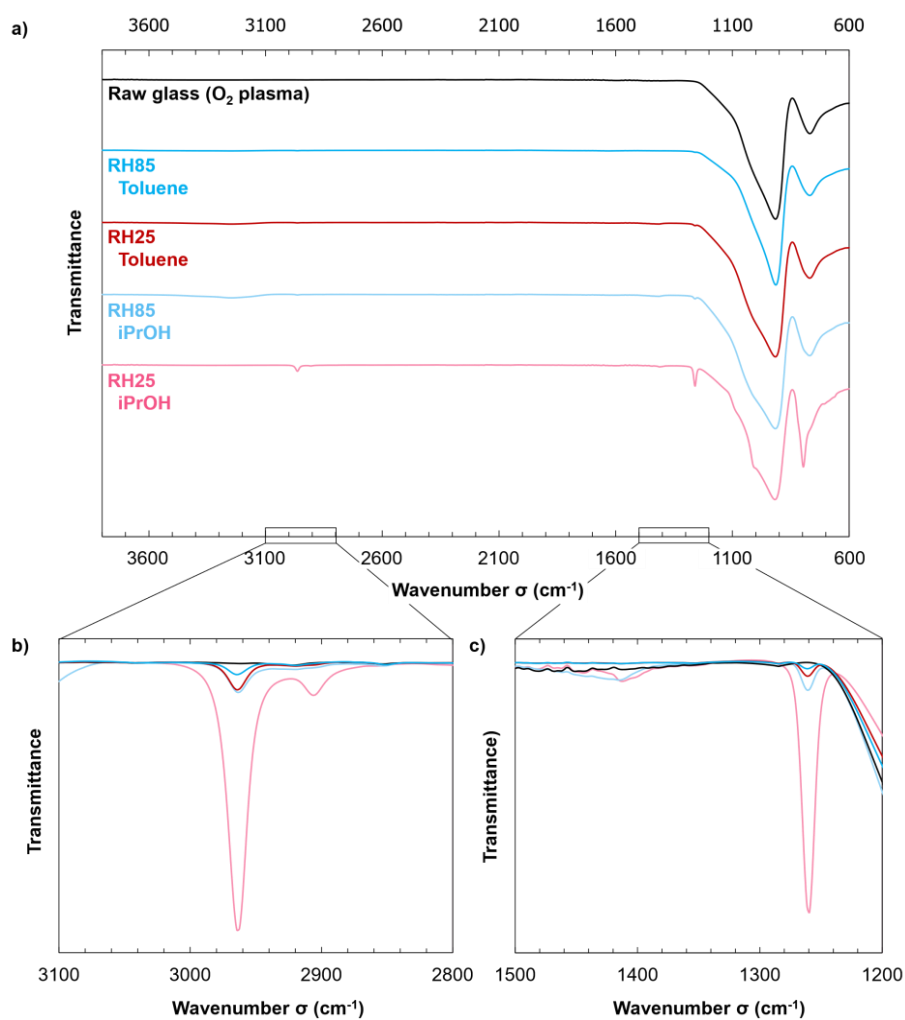

*Figure SI.C.1.* ATR-FTIR spectra of raw glass (O$_2$ plasma cleaned) and of DMDMOS-based coatings made under RH = 25 % (red curve) or RH = 85 % (blue curve), with a H$_2$O+iPrOH rinsing (lighter curve) or a H$_2$O+iPrOH+toluene rinsing (darker curve). These spectra were obtained after correction of the baseline with OPUS correction tool.

## D – XPS measurements

The surface chemical composition of raw glass and of DMDMOS coated glass has been studied by XPS. The samples have been made under RH = 25 or 85 % (respectively identified as RH25 and RH85) and rinsed with H$_2$O+iPrOH+toluene. The Si 2p spectrum of raw glass (Figure SI.D.1.a) was fitted by one feature at 103.5 eV for SiO$_2$ bonding [66,67]. After coating with DMDMOS, the Si 2p signal widened, highlighting the presence of another peak at 102.2 eV (Figure SI.D.1.a). This binding energy was attributed to PDMS molecules [67–70], which confirmed the efficiency of the grafting for both RH. Another clear evidence of PDMS grafting was found on C 1s spectra (Figure SI.D.1.b). After coating, the peaks related to contamination carbons (C=O, C–O, C–C) were replaced by a single peak located at 284.8 eV, specific of Si–C bonding in PDMS molecules [68,69].

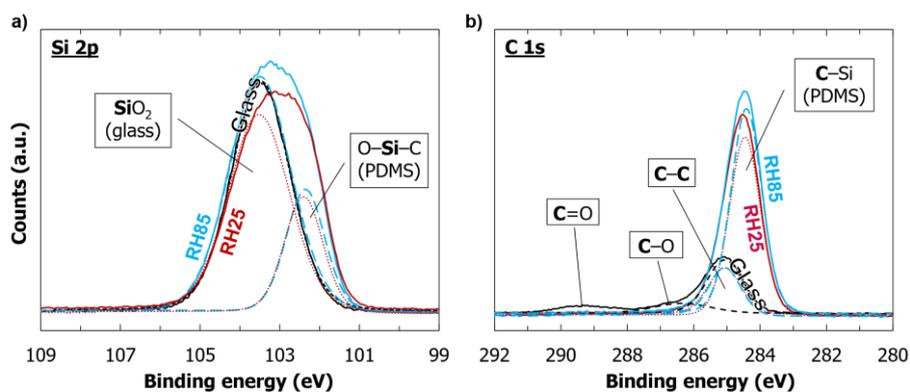

*Figure SI.D.1.* **a)** Si 2p and **b)** C 1s XPS spectra of raw glass (black curve) and with DMDMOS-based coatings made under RH = 25 % (red curve) or RH = 85 % (blue curve), after a H$_2$O+iPrOH+toluene rinsing. The fitting curves are represented with dash lines: black small dashed lines for raw glass, red dotted lines for RH25 sample and blue dashed lines for RH85 sample. In figure a), the SiO$_2$ fitting curves are overlapping the raw glass and the RH85 experimental spectra.